\pdfoutput=1

\documentclass[11pt,twoside,a4paper,cmspaper,final,collab]{cms-tdr}
\def\svnVersion{46267:46271M}\def\cmsCernNo{2011-024}\def\cmsCernDate{\today}\def\cmsMessage{Submitted to the Journal of High Energy Physics}

\begin{document}\cmsNoteHeader{EWK-10-006}

\hyphenation{had-ron-i-za-tion}
\hyphenation{cal-or-i-me-ter}
\hyphenation{de-vices}
\RCS$Revision: 46203 $
\RCS$HeadURL: svn+ssh://alverson@svn.cern.ch/reps/tdr2/papers/EWK-10-006/trunk/EWK-10-006.tex $
\RCS$Id: EWK-10-006.tex 46203 2011-03-17 14:35:54Z mpioppi $

\newcommand{\wlnu}{\mathrm{W} \rightarrow \ell\nu}
\newcommand{\wmunu}{\mathrm{W} \rightarrow \mu\nu}
\newcommand{\wtaunu}{\mathrm{W} \rightarrow \tau\nu}
\newcommand{\wenu}{\mathrm{W} \rightarrow \mathrm{e}\nu}
\newcommand{\zll}{ \mathrm{Z}/\gamma^*\rightarrow \ell^+ \ell^-}
\newcommand{\zmumu}{ \mathrm{Z}/\gamma^*\rightarrow \mu^+ \mu^- }
\newcommand{\zee}{ \mathrm{Z}/\gamma^*\rightarrow \mathrm{e}^+ \mathrm{e}^- }
\newcommand{\ztautau}{ \mathrm{Z}/\gamma^*\rightarrow \tau^+ \tau^- }

\newcommand{\myttbar}{ \mathrm{t}\bar{\mathrm{t}}}

\cmsNoteHeader{EWK-10-006} 
\title{Measurement of the Lepton Charge Asymmetry in Inclusive $\mathrm{W}$ Production in $\mathrm{pp}$ Collisions at $\sqrt{s} =
7\,\mathrm{Te\hspace{-.08em}V} $ }

\address[neu]{Northeastern University}
\address[fnal]{Fermilab}
\address[cern]{CERN}
\author[cern]{The CMS Collaboration}

\date{\today}

\abstract{

A measurement of the lepton charge asymmetry in
inclusive $\mathrm{pp}\rightarrow\mathrm{W } X$ production
at $\sqrt{s}= 7\,\mathrm{Te\hspace{-.08em}V} $ is presented based on
data
recorded by the
CMS detector at the LHC and corresponding to an integrated luminosity
of $36\ \mathrm{pb}^{-1}$.
This high precision measurement of the lepton charge asymmetry,
performed in both the $\mathrm{W} \rightarrow \mathrm{e} \nu$ and
$\mathrm{W} \rightarrow \mu \nu$ channels,
provides new insights into parton distribution functions.
}

\hypersetup{%
pdfauthor={CMS Collaboration},%
pdftitle={Measurement of the lepton charge asymmetry in inclusive W production in pp collisions at sqrt(s) = 7 TeV},%
pdfsubject={CMS},%
pdfkeywords={CMS, physics, electro-weak}}

\maketitle 

\section{Introduction}
In $\Pp\Pp$ collisions, $\PW$ bosons are produced primarily via the
 processes
$\mathrm{u}\mathrm{\bar{d}}\rightarrow\PWp$ and $\mathrm{d}\mathrm{\bar{u}}\rightarrow\PWm$.
The first quark is a valence quark from one of the protons, and the second one is a sea antiquark from
the other proton. Due to the presence of two valence $\mathrm{u}$ quarks in the proton,
there is an overall excess of $\PWp$ over $\PWm$ bosons.
The inclusive ratio of cross sections for $\PWp$ and $\PWm$ bosons production at the Large
Hadron Collider~(LHC)
was measured
to be $1.43 \pm 0.05$ by the Compact Muon Solenoid~(CMS) experiment~\cite{CMS:WZ}
and is in agreement with predictions of the Standard Model~(SM) based on various
parton distribution functions~({\sc PDF}s)~\cite{MSTW:0901, CTEQ:1007}.
 Measurement of this production asymmetry
between $\PWp$ and $\PWm$ bosons as a function of boson rapidity
can provide new insights on the $\mathrm{u}/\mathrm{d}$ ratio and the sea antiquark densities in 
the ranges of the Bj\"{o}rken parameter $x$~\cite{BJORKEN} probed in $\Pp\Pp$
collisions at $\sqrt{s}=7\TeV$.
However, due to the presence of neutrinos in leptonic $\PW$ decays the boson rapidity is not directly accessible.
The experimentally accessible quantity 
is  the
lepton charge asymmetry, defined to be

\def\difsig{\ensuremath{\mathrm{d}\sigma/\mathrm{d}\eta}}

$$
  \mathcal{A}(\eta) = \frac{\difsig(\PWp\rightarrow\ell^+\nu)-\difsig(\PWm\rightarrow\ell^-\bar{\nu})}
   {\difsig(\PWp\rightarrow\ell^+\nu)+\difsig(\PWm\rightarrow\ell^-\bar{\nu})},
$$

where $\ell$ is the daughter charged lepton, $\eta$ is the charged lepton
pseudorapidity in the
CMS lab frame ($\eta= -\ln{[\tan{(\frac{\theta}{2})}]}$ where $\theta$ is the polar angle), and
$\mathrm{d}\sigma/\mathrm{d}\eta$ is the differential cross section for charged
leptons from $\PW$ boson decays.
The lepton charge
asymmetry can be used to test SM predictions with high
precision.
Due to the $\mathrm{V}-\mathrm{A}$ structure of the $\PW$ boson couplings to fermions,
theoretical predictions of the charge asymmetry
depend on the transverse momentum~($\pt$) threshold applied on the daughter leptons.
For this reason, we measure $\mathcal{A}(\eta)$ for two different charged lepton $\pt$~($\pt^{\ell}$) thresholds,
25\GeVc and 30\GeVc.

The lepton charge asymmetry and the $\PW$ charge
asymmetry have been studied in  $\mathrm{p}\bar{\mathrm{p}}$ collisions by both
the CDF and D0 experiments at the Fermilab Tevatron Collider~\cite{CDF:wasym, D0:asym}.
Current predictions for the lepton charge
asymmetry at the LHC based on different {\sc PDF} models do not
agree well with each other.
A high precision measurement of this asymmetry
at the LHC
can contribute to the improvement of  the knowledge of {\sc PDF}s.
The 
measurement of the muon charge asymmetry at the LHC with a dataset
corresponding to an integrated luminosity of about 31 $\mathrm{pb}^{-1}$ was
reported recently by the ATLAS experiment~\cite{ATLAS:asym}. In this Letter,
we present a measurement of the lepton charge asymmetry in both
$\wenu$
and $\wmunu$ final states using a
dataset corresponding to an integrated luminosity of 36
$\mathrm{pb}^{-1}$
collected by the CMS detector in March--November 2010.

\section{The CMS Detector}
A detailed description of the CMS experiment can be found
elsewhere~\cite{refCMS}.
The central feature of the CMS apparatus is a
superconducting solenoid, of 6~m internal diameter, 13~m in length,
providing an axial field of
3.8~T. Within the field volume are the silicon pixel and strip tracker,
the crystal electromagnetic calorimeter (ECAL) and the brass/scintillator
hadron calorimeter (HCAL). Muons are measured in gas-ionization detectors
embedded in the steel return yoke of the solenoid.
The most relevant sub-detectors for this measurement are the ECAL,
the muon system, and the tracking system. The electromagnetic calorimeter consists
of nearly 76\,000 lead tungstate crystals which provide coverage in pseudorapidity $|\eta| <$ 1.479  in the
 barrel region and 1.479 $< |\eta| <$ 3.0 in two endcap regions.
A preshower detector consisting of two planes of silicon sensors interleaved with a total of 3$\,X_0$ of lead is located in front of the ECAL endcaps.
The ECAL has an ultimate energy resolution of better than 0.5\% for
unconverted photons with transverse energies above 100\GeV.
 The electron energy
resolution is 3\% or better for the range of electron energies relevant for
this analysis.
Muons are measured in the pseudorapidity range $|\eta| <$ 2.4,
with detection planes made of three technologies: drift tubes, cathode strip chambers, and resistive plate chambers. Matching the muons to the tracks measured in the silicon tracker results in a transverse momentum resolution of about
2\% in the relevant muon $\pt$ range.

CMS uses a right-handed coordinate system, with the origin at the nominal interaction point, the $x$-axis pointing to the center of the LHC, the $y$-axis pointing up (perpendicular to the LHC plane), and the $z$-axis along the anticlockwise-beam direction. The polar angle, $\theta$, is measured from the positive $z$-axis and the azimuthal angle, $\phi$, is measured in the $x$-$y$ plane.

\section{Analysis Method and Simulation}

The  $\wlnu$ candidates are characterized by a
high-$\pt$
lepton accompanied by missing transverse energy $\Em_{\mathrm{T}}$,  due
to the escaping neutrino.
Experimentally,  $\Em_{\mathrm{T}}$ is determined as
 the negative vector sum of the
 transverse momenta of
all particles reconstructed using a particle flow algorithm~\cite{pfmet}.
The $\wenu$ and
 $\wmunu$ candidates used in this analysis were
collected using a set of inclusive
single-electron and single-muon triggers which did not include $\Em_{\mathrm{T}}$
requirements.
Other physics processes, such as  multijet and photon+jet
production~(QCD background),
Drell--Yan~($\zll$) production, $\wtaunu$ production~(EWK background),
and top quark pair ($\myttbar$) production can produce
 high-$\pt$ electron/muon candidates
and mimic
$\PW$ candidates. Furthermore,
cosmic ray muons can produce fake
$\wmunu$ candidates.
The muon measurement relies on the  muon detector, inner tracking and
calorimeters
to form a
robust isolation
variable
to separate signal from background.
This method is not applicable to the electron measurement because electron candidates are accompanied by significant electromagnetic radiation, reducing the power of the isolation in extracting signal from background.
Instead
the electron selection relies on
the calorimeter system and uses the $\Em_{\mathrm{T}}$ to
separate signal from background. These
two measurements are largely independent and cross-check each other.

Monte Carlo~(MC) simulation samples have been
used to develop analysis techniques and estimate some of the background contributions.
The next-to-leading order~(NLO) MC
 simulations based on the {\sc POWHEG} event generator~\cite{POWHEG}
 interfaced  with the CT10 PDF model~\cite{CTEQ:1007} are primarily used.
The  QCD multijet background
is generated with the {\sc PYTHIA} event generator~\cite{PYTHIA6} interfaced with
the CTEQ6L PDF model~\cite{CTEQ6L}.
The $\myttbar$ background is generated using both {\sc PYTHIA} and
{\sc MC@NLO}~\cite{MC@NLO} event generators.
Other {\sc PYTHIA}-based MC simulations
are used to cross-check MC predictions and help assigning experimental
systematic uncertainties.  The {\sc PYTHIA}-based MC samples are normalized using NLO cross sections except for the QCD background samples.
All generated events are passed
through the  CMS detector simulation using  {\sc GEANT4}~\cite{GEANT4}
and then processed using a reconstruction sequence identical to that used for
 collision data.
Pile-up, which consists of the presence
of secondary minimum bias interactions in addition to the primary hard
interaction in an event,
is significant in the data, where
an average of 2--3 vertices are found.
The analysis methods used in both electron and muon channels are
insensitive to the effect of pile-up.

The selection criteria for electron/muon reconstruction and identification
are almost identical to those used in the $\PW$ and
$\mathrm{Z}$ cross section measurements~\cite{CMS:WZ}. A brief summary is given here
for completeness.

\section{Electron Reconstruction and $\wenu$ Signal Extraction}
Electrons are identified as clusters of energy deposited in the ECAL fiducial
volume matched to tracks from the inner silicon tracker~({\it silicon tracks}).
The silicon tracks are
reconstructed using a Gaussian-Sum-Filter~(GSF)
algorithm~\cite{GSF} that takes into account possible energy loss
due to bremsstrahlung in the tracker layers. Particles misidentified  as
electrons are suppressed by requiring that the shower shape of the
ECAL cluster
be  consistent with an electron candidate,
and that the $\eta$ and $\phi$ coordinates
of the track trajectory extrapolated to the ECAL match the $\eta$ and $\phi$
coordinates of the ECAL cluster.
Furthermore, electrons from $\PW$ decay are isolated from other activity
in the event.
We therefore require that little transverse energy be observed in the
ECAL, HCAL, and silicon tracking system within a cone
$\Delta R <$ 0.3 around the electron direction, where $\Delta R=\sqrt{(\Delta\phi)^2+(\Delta\eta)^2}$ and where calorimeter energy deposits and the track
associated with the electron candidate are excluded.
 Due to
the substantial  amount of material in front of the ECAL detector,
a large fraction of electrons radiate photons. The resulting photons
may convert close to the original electron trajectory, leading to
a sizable charge misidentification rate~($w$).
The true charge asymmetry, $\mathcal{A}^{\textrm{true}}$,  is diluted
 due to charge misidentification resulting in an observed asymmetry, $\mathcal{A}^{\textrm{obs}}= \mathcal{A}^{\textrm{true}}(1-2w).$
Three different methods are used to
determine the electron charge. First, the electron charge
is  determined by the signed curvature of the
associated GSF track. Second, the charge is determined from the
associated trajectory reconstructed in the silicon tracker
using a Kalman Filter algorithm~\cite{KF}. Third,
the electron charge is determined based on the azimuthal angle between
the vector joining
the nominal interaction point and the ECAL cluster position and the vector
joining the nominal interaction point and innermost hit of the GSF track. It is required that  all
three charge determinations from these methods agree.
This procedure significantly reduces the charge misidentification rate
to 0.1\% in the ECAL barrel and to 0.4\%
in the ECAL endcaps.
The $\wenu$ candidates are selected by further
requiring electrons to have  $\pt >$ 25\GeVc, $|\eta| <$ 2.4,  and
to be associated with one of the electron
trigger candidates used to select the electron dataset. The Drell--Yan
and $\myttbar$ backgrounds are  suppressed
by rejecting events that contain a second isolated
electron or muon with $\pt >$ 15\GeVc and $|\eta| <$ 2.4.
According to MC simulations,
the data sample of selected electrons
consists of
 about 28\% QCD background events,
about
 6.5\%  EWK background
events,
and about 0.2\% $\myttbar$ background events.
The events passing the above selection criteria are divided into six bins of electron
pseudorapidity ($|\eta^\textrm{e}|$): [0.0,\ 0.4],\ [0.4,\ 0.8],\ [0.8,\ 1.2],\
[1.2,\ 1.4],\ [1.6,\ 2.0],\ and\  [2.0,\ 2.4],
 for the measurement of the electron charge asymmetry.~(The fourth bin is
reduced to a width of 0.2 in order to
exclude the transition region between the ECAL barrel and endcaps.)

\begin{figure}[tb]
  \begin{center}
    \includegraphics[width=0.95\columnwidth]{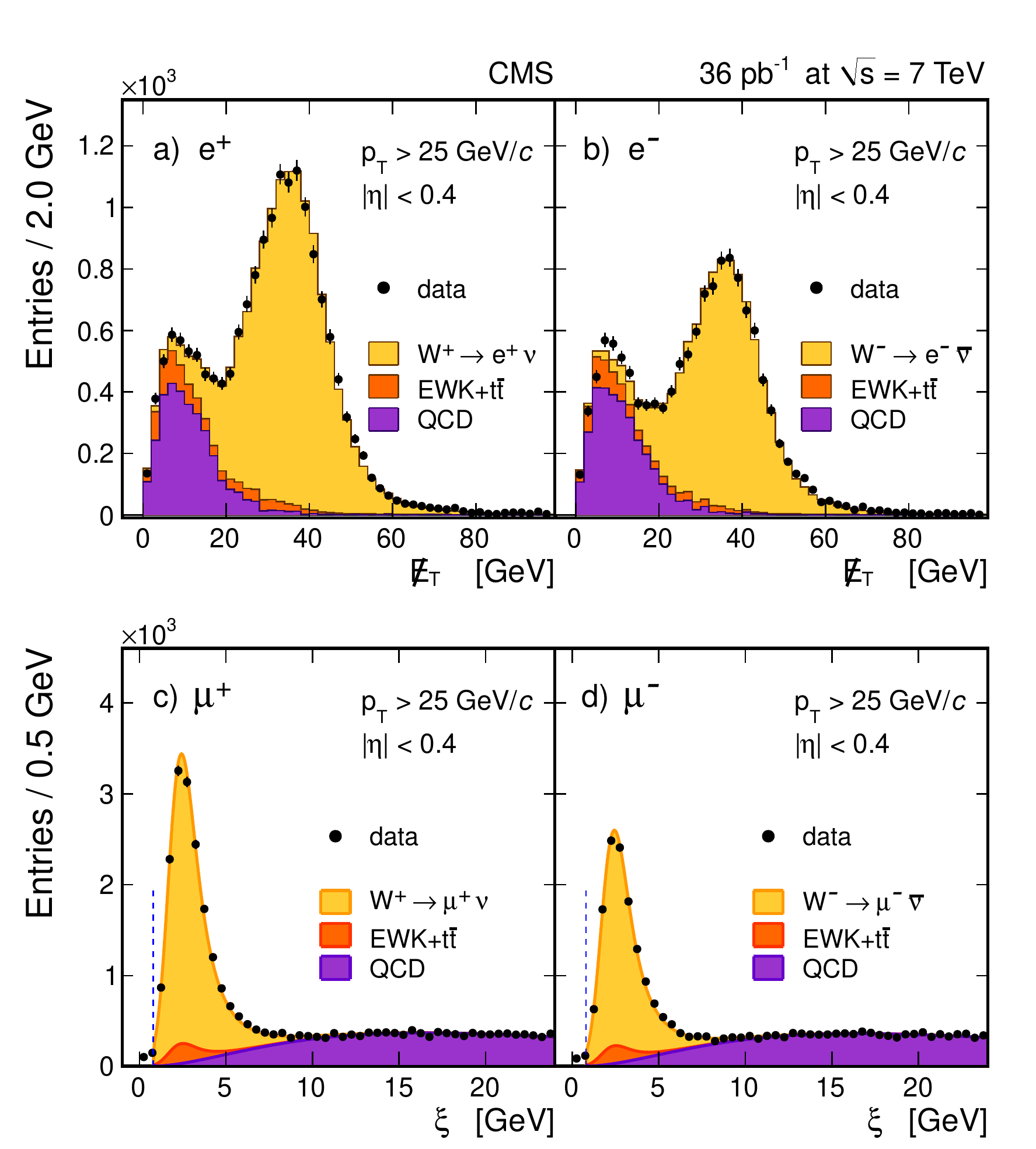}
    \hspace{1cm}
    \caption{Signal fit to data $\Em_{\mathrm{T}}$  distributions for electrons,
a) $\PWp \rightarrow \mathrm{e}^+\nu$,
    b) $\PWm\rightarrow \mathrm{e}^-\bar{\nu}$, and  fit  to
$\xi$ distributions
for muons,  c) $\PWp \rightarrow \mu^+\nu$, and d) $\PWm \rightarrow \mu^-\bar{\nu}$.
Only the results for
the first pseudorapidity bin ($|\eta| <$ 0.4) are shown. In Figs.~c) and d),
only events on the right of the dashed vertical line are included in the
fit. The QCD (multijet and photon+jet production) shape is determined directly from data.}
    \label{fig1}
  \end{center}
\end{figure}

A binned extended maximum
likelihood fit is performed
over the $\Em_{\mathrm{T}}$ distribution to estimate the $\wenu$ signal
yield for electrons~($N^-$) and positrons~($N^+$) in each pseudorapidity bin.
The signal $\Em_{\mathrm{T}}$ shape is derived
from MC simulations with an event-by-event correction to account for energy
scale and resolution differences between data and MC based on
 the hadronic recoil energy distributions in $\zee$ events selected
from data~\cite{RECOIL}.
 The shape of the QCD background is
determined, for each charge, from a signal-free control sample obtained by
inverting a subset of the electron identification criteria.
The $\Em_{\mathrm{T}}$ shapes for
other backgrounds such as the Drell--Yan process,
 $\myttbar$,
 and $\wtaunu$ are taken from MC simulations with
a fixed  normalization relative to the $\wenu$
yields. The normalization factors are from
the predicted values of the cross sections at NLO.
The yield of the  QCD  background and the
 yield of the $\wenu$ signal are free parameters
in the fit. The results of the fits to the data
are shown for the first pseudorapidity bin in
Figs.~\ref{fig1}~(a) and~\ref{fig1}~(b).
The charge asymmetry is
obtained from
$(N^+-N^-)/(N^++N^-)$.

\section{Muon Reconstruction and $\wmunu$ Signal Extraction}
Muon candidates are reconstructed using two different algorithms: one starts
from inner silicon tracks
 and
requires a minimum number of matching hits in the muon chambers,
and
the other finds tracks in the muon system and matches them to
silicon tracks.
A global track fit including both the silicon track hits
and muon chamber hits
is performed to improve the quality of the
reconstructed muon candidates.
 The muon candidate is not required
in this analysis to be isolated from other event activity, a
difference in muon selection with respect to~\cite{CMS:WZ}, in order to avoid
bias in the signal extraction fit described below.
The muon charge is identified from the signed curvature of
 the associated
 silicon track.
A selection on
the silicon track distance of closest approach to the beam spot,
$|d_{xy}|<$ 0.2 $ \mathrm{cm}$,  is applied to reduce the cosmic ray background.
The remaining cosmic ray background yield is estimated by
normalizing the $|d_{xy}|$ distribution derived from cosmic ray muon data to the
large $|d_{xy}|$ region~(1.0 $< |d_{xy}| <$ 5.0 $\mathrm{cm}$) in the data sample.
The
 estimated
cosmic ray background contamination is
  about $10^{-5}$
 of the expected $W\rightarrow \mu\nu $ signal
yield
and
is neglected.
The
$\wmunu$ candidates are selected by further
requiring the muon $\pt$ to be greater than  25\GeVc,
$|\eta| <$ 2.1,  and that the candidate  matches
one of the muon trigger candidates.
The Drell--Yan background
is   suppressed by rejecting events which contain a
second isolated muon
with $\pt >$ 15\GeVc, $|\eta| <$ 2.4, and
passing the above muon quality selections.
The events which passed the above selection criteria are divided into six bins of muon
pseudorapidity ($|\eta^{\mu}|$):
[0.0,\ 0.4],\ [0.4,\ 0.8],\ [0.8,\ 1.2],\
[1.2,\ 1.5],\ [1.5,\ 1.8],\ and\  [1.8,\ 2.1].

The $\wmunu$ signal estimation is done by fitting
 the distribution of an isolation variable
$\xi=\Sigma(E_T)$
defined as the scalar sum of the transverse momenta
of silicon tracks (excluding the muon candidate)
and  energy deposits in both ECAL and HCAL in a cone $\Delta R <$ 0.3 around the muon direction.
The calorimeter energy deposit associated with the muon candidate is rescaled
to ensure a uniform transverse energy response as a function of polar angle.
With this correction the shape of the
signal $\xi$ distribution
becomes independent of pseudorapidity, peaking at a constant $E_{\mathrm{T}}$
value of about 2.5\GeV.
The correction was obtained from muons in the $\zmumu$ MC sample and checked with real muons from the $\zmumu$ data sample.
The shape of the $\xi$ distribution for muons from  $\wmunu$
is parametrized as a Landau distribution
convolved with a Gaussian resolution function.
The tail of the Landau distribution is modified to be an exponential
function.
The $\wmunu$ signal, the Drell--Yan background,
the $\myttbar$ background,  and the $\wtaunu$ background
have been shown in MC simulations  to have $\xi$ distributions that are
identical and
lepton-charge independent.
The charge independence has been confirmed in studies of $\zmumu$ data.
In the final fit the shape parameters for the signal $\xi$
distribution are fixed using
 $\zmumu$ data,  except for  the exponential
tail parameters which are fixed using  $\zmumu$ MC
simulations due to limited sample of $\zmumu$ in data.
The QCD background is parametrized by an empirical function
$\xi^{\alpha}\cdot e^{\beta \sqrt{\xi}}$.
The QCD parametrization and the value of  background parameter $\alpha$
 is determined directly from data using a QCD background control sample obtained
by selecting events with
large impact parameter significance and little $\Em_{\mathrm{T}}$
in the event.
The background parameter $\beta$ is allowed to float.
The QCD yield is 
determined separately
for negative and positive charges.
The region used for signal estimation is chosen to be
 0 $< \xi <$ 25\GeV. From MC studies,
the expected
QCD multijet, EWK, and $\myttbar$ backgrounds are
about 13.0\%, 6.9\%, and 0.3\%, respectively
 within the interval
0 $< \xi <$ 10\GeV,  in which most of $\wmunu$ signal candidates are found.

An unbinned extended maximum likelihood
fit to the $\xi$ distribution
is performed simultaneously on the $\PWp \rightarrow \mu^+\nu$ and
$\PWm \rightarrow \mu^-\bar{\nu}$ candidates to determine
 the total $\wmunu$ signal yield and the charge asymmetry in
each pseudorapidity bin.
Background events from $\zmumu$ are selected
if one of the daughter muons is outside of the detector acceptance.
This part of the $\zmumu$ background is
normalized
to the observed $\zmumu$
 events in data obtained by inverting the Drell--Yan veto selection.
The acceptance ratio for one versus two muons from $\zmumu$ events
is estimated with MC simulations.
There are  small contributions to the
 $\zmumu$ background with both muons
 staying
within the detector coverage but one of them
failing  reconstruction, muon identification, or
 isolation requirement. This background contribution
 is estimated directly from data. The
$\ztautau$ background is normalized to the
estimated $\zmumu$
background with ratios determined from MC simulations.
 The $\wtaunu$ background  is
normalized to the $\wmunu$ signal yield in the data using the
ratio between $\wtaunu$  background and $\wmunu$ signal
yield determined from MC simulations. The $\myttbar$ background is estimated
directly from 
MC simulations with cross sections normalized to the predicted value at NLO.
The EWK and $\myttbar$ backgrounds
are estimated
for the $\PWp \rightarrow \mu^+\nu$ and
$\PWm \rightarrow \mu^-\bar{\nu}$ candidates separately.
The results of the fits to the data
are shown for the first pseudorapidity bin in
Figs.~\ref{fig1}~(c) and~\ref{fig1}~(d).
Only the region $\xi >$ 0.8\GeV is included in the fit because
the small region  $\xi <$ 0.8\GeV exhibits a complex shape for both
signal and QCD background due to the zero-suppressed readout of the CMS
calorimeters.  However,
events in this region are included to determine  the
$\wmunu$ signal
yield and charge asymmetry, as  the number of QCD background events in this
region is
negligible, confirmed by a Monte Carlo simulation study.

\section{Systematic Uncertainties}
The systematic uncertainties considered for both the electron and muon channels are mainly due to the
lepton charge misidentification rate, possible efficiency differences between the $\ell^+$
and $\ell^-$,
lepton momentum~(energy) scale and
resolution,
and signal estimation.

 The electron charge misidentification rate is measured in data
using the $\zee$ data sample to be within 0.1--0.4\%, increasing with electron pseudorapidity.
The measured electron charge asymmetry is corrected for the charge
misidentification rate.
The statistical error on the electron
charge misidentification rate is taken as the systematic
uncertainty.
The muon charge misidentification rate is studied using $\wmunu$
MC simulations and is estimated to be at the level of
$10^{-5}$.
The muon charge misidentification rate is further studied  with a cosmic ray muon data
sample in which one cosmic ray muon passes through the center of the CMS detector
and
 is  reconstructed as two muon candidates with opposite charge~\cite{COSMICS}.
A total of $16\,422$ cosmic ray muon events with at least two muon candidates
are selected,
and no event is found to
have a same sign muon pair. This constrains the muon
charge misidentification rate to be less than $10^{-4}$. Charge misidentification therefore
has a negligible
effect on the measured muon charge asymmetry.

The efficiency difference between $\ell^+$ and $\ell^-$
can result in a bias on the measured charge asymmetry.
The total lepton efficiency (including lepton reconstruction, identification,
and trigger efficiencies) in each pseudorapidity bin is measured
using the $\zll$ data for $\ell^+$ and
$\ell^-$, respectively. The efficiency ratio is calculated and
found to be
 consistent with unity within the statistical uncertainty. 
No correction is made to the observed charge asymmetry. However,
the statistical errors on the efficiency ratios are treated
as systematic uncertainties.
The inclusive
electron efficiency ratio is found to be $1.007\pm0.014$, and the error on this
ratio is used to determine the systematic uncertainty for the electron
channel.
Within the statistical uncertainty the muon efficiency ratio is also constant across the
 pseudorapidity coverage. However,
due to a small known
charge/pseudorapidity-dependent muon momentum scale bias, the statistical uncertainty on the
bin-by-bin efficiency ratio is used.
This is
the dominant systematic uncertainty for
both the electron and muon channels.

In order to compare our results directly
to theoretical predictions, the measured charge asymmetry in the electron (muon) channel is corrected
for lepton energy~(momentum)
bias and resolution
effects.
The lepton scale and resolution are  determined directly from  the $\zll$
data and are used to smear MC lepton $\pt^{\ell}$ at the generator level. The correction to
the measured
charge asymmetry  is estimated in each pseudorapidity bin
by comparing the   charge asymmetry as determined in the MC
with the resulting
asymmetry after smearing.
The
 uncertainties on the energy~(momentum) scale and
resolutions are taken as sources for
systematic uncertainties. The electron energy scale bias is studied using $\zee$
 data and determined to be
within 1\%. The  charge-dependent muon momentum
scale bias is also studied using $\zmumu$ data sample and
 determined to be within 1\%.
 This energy~(momentum) scale bias dominates the systematic uncertainty due to
lepton energy~(momentum) scale and resolution.
The impact of the QED final-state
 radiation~(FSR)
 on the lepton charge asymmetry is also studied using {\sc POWHEG} MC samples and a
reduction  of the asymmetry at the level of 0.1--0.2\% is found.
The charge asymmetry is corrected for the FSR effect and
the full correction is taken as additional
systematic uncertainty which is summed in quadrature with
the lepton energy~(momentum) scale  systematic uncertainty.

For the electron channel,
the dominant systematic uncertainty on the signal and background estimation is from the
modeling of the $\Em_{\mathrm{T}}$ shape for the signal.
Others,  such as the uncertainties on the
background cross sections and QCD background shape modeling,
also contribute. The systematic uncertainty due to the QCD
background shape is studied by using different QCD background control regions
to derive the QCD background $\Em_{\mathrm{T}}$ shape.
For the muon channel,
the systematic sources for signal and background estimation are due to
signal and background parametrization, uncertainty in estimating
Drell--Yan and $\myttbar$ background yields, and the
$\wtaunu$ background to $\wmunu$ signal ratio.
The largest contribution is from
the estimation of Drell--Yan
background
which is dominated
by the limited number
 of Drell--Yan dimuon events and  variations
of the acceptance ratio for one versus two muons from $\zmumu$ events.
This acceptance ratio depends on both PDF models and the $\mathrm{Z}$ boson
$\pt$ spectrum. The {\sc PYTHIA} and {\sc POWHEG} MC samples are used
to estimate the ratios, and differences between these two MC simulations~(at the level of 5--10\%) are treated as systematic uncertainties.
The $\myttbar$
background yield is varied by 18.6\% to reflect
the theoretical uncertainty on the $\myttbar$ cross section.
The ratio of
the $\wtaunu$
background to the $\wmunu$ signal and ratio of
 the $\ztautau$
background to the $\zmumu$ background
are estimated using both {\sc POWHEG} and {\sc PYTHIA} MC simulations.
The differences between these two MC simulations are taken  as sources of
 systematic uncertainties.
The robustness of the  signal parametrization is studied using pseudo-experiments,
whose pseudo-data are
obtained by sampling fully simulated MC events according to Poisson distributions with means set to
the measured signal and background yields in data.
 A small bias
on the measured  charge asymmetry at the level of 0.1--0.2\% is
found and taken as additional systematic uncertainty
on the signal estimation.
\begin{table*}[tb]
    \caption{Summary of the systematic uncertainties. All values are in percent.
}
    \label{tab1}
    \begin{center}
{\footnotesize

      \begin{tabular*}{\textwidth}{@{\extracolsep{\fill}}  l|rrrrrr|rrrrrr} \hline\hline

\multicolumn{13}{c}{\rule[-1.8mm]{0mm}{4mm} $p_{\mathrm{T}}^{\ell} >$ 25\GeVc \rule[-1.8mm]{0mm}{6.0mm}}\\
\cline{1-13}

&\multicolumn{6}{c|}{Electron Channel}&   \multicolumn{6}{c}{Muon Channel}\\

$|\eta|$ bin & [0.0, & [0.4, & [0.8, & [1.2, & [1.6, & [2.0, &
[0.0, & [0.4, & [0.8, & [1.2, & [1.5, & [1.8, \\
& 0.4]& 0.8]& 1.2]& 1.4]& 2.0]& 2.4]&
0.4]& 0.8]& 1.2]& 1.5]& 1.8]& 2.1]\\\hline

Charge Misident.   & 0.02& 0.03& 0.03& 0.08& 0.09& 0.10& 0& 0& 0& 0& 0& 0\\
Eff. Ratio         & 0.70& 0.70& 0.70& 0.70& 0.70& 0.70& 0.59& 0.39& 0.92& 0.72& 0.81& 1.17\\
$ e/\mu$  Scale    & 0.11& 0.09& 0.19& 0.47& 0.40& 0.45& 0.50& 0.48& 0.50& 0.48& 0.50& 0.42\\
Sig. \& Bkg. Estim.& 0.16& 0.19& 0.26& 0.33& 0.25& 0.25& 0.23& 0.29& 0.34& 0.40& 0.53& 0.58\\\hline
Total              & 0.73& 0.73& 0.77& 0.90& 0.85 &0.87 & 0.80& 0.68& 1.10& 0.95& 1.08& 1.37\\\hline\hline

\multicolumn{13}{c}{ \rule[-1.8mm]{0mm}{5.5mm} $p_{\mathrm{T}}^{\ell} >$ 30\GeVc }\\\cline{1-13}

&\multicolumn{6}{c|}{Electron Channel}&   \multicolumn{6}{c}{Muon Channel}\\

$|\eta|$ bin & [0.0, & [0.4, & [0.8, & [1.2, & [1.6, & [2.0, &
[0.0, & [0.4, & [0.8, & [1.2, & [1.5, & [1.8, \\
& 0.4]& 0.8]& 1.2]& 1.4]& 2.0]& 2.4]&
0.4]& 0.8]& 1.2]& 1.5]& 1.8]& 2.1]\\\hline

Charge Misident. & 0.02& 0.02& 0.03& 0.07& 0.08& 0.10& 0& 0& 0& 0& 0& 0\\
Eff. Ratio       & 0.70& 0.70& 0.70& 0.70& 0.70& 0.70& 0.59& 0.39& 0.93& 0.72& 0.82& 1.18\\
$ e/ \mu$  Scale & 0.07& 0.17& 0.26& 0.46& 0.53& 0.55& 0.80& 0.78& 0.83& 0.81& 0.73& 0.77\\
Sig. \& Bkg. Estim.  & 0.16& 0.19& 0.26& 0.33& 0.25& 0.25& 0.20& 0.20& 0.27& 0.35& 0.51& 0.56\\\hline
Total            & 0.72& 0.75& 0.79& 0.91& 0.92& 0.93& 1.01& 0.90& 1.27& 1.14& 1.21& 1.52\\\hline\hline

      \end{tabular*}

}
    \end{center}
  \end{table*}

\section{Results and Conclusions}

Table~\ref{tab1} summarizes systematic uncertainties for both the
electron
and muon channels.
The measured charge asymmetry results are summarized in
Table~\ref{tab2}
with both statistical
and systematic uncertainties shown.
The measurements have been repeated with a lepton   $\pt^{\ell} >$ 30\GeVc. This
requirement
selects a subset of events with lepton pseudorapidity closer to the
$\PW$ boson rapidity and enables us to test PDF predictions in a more
constrained region of phase space.
The electron and muon measurements are in agreement with each other.
The experimental results are compared to theoretical predictions
obtained using {\sc RESBOS}~\cite{RES1,RES2,RES3} and {\sc MCFM}~\cite{MCFM} generators interfaced with
CT10W PDF model~\cite{CTEQ:1007}.
The {\sc RESBOS} generator performs a resummation at the next-to-next-to-leading
logarithmic order and gives a
more realistic
description of the $\PW$ boson $\pt$ spectrum
than a fixed-order calculation.

\begin{figure}[tb]
  \begin{center}
    \includegraphics[width=0.75\columnwidth]{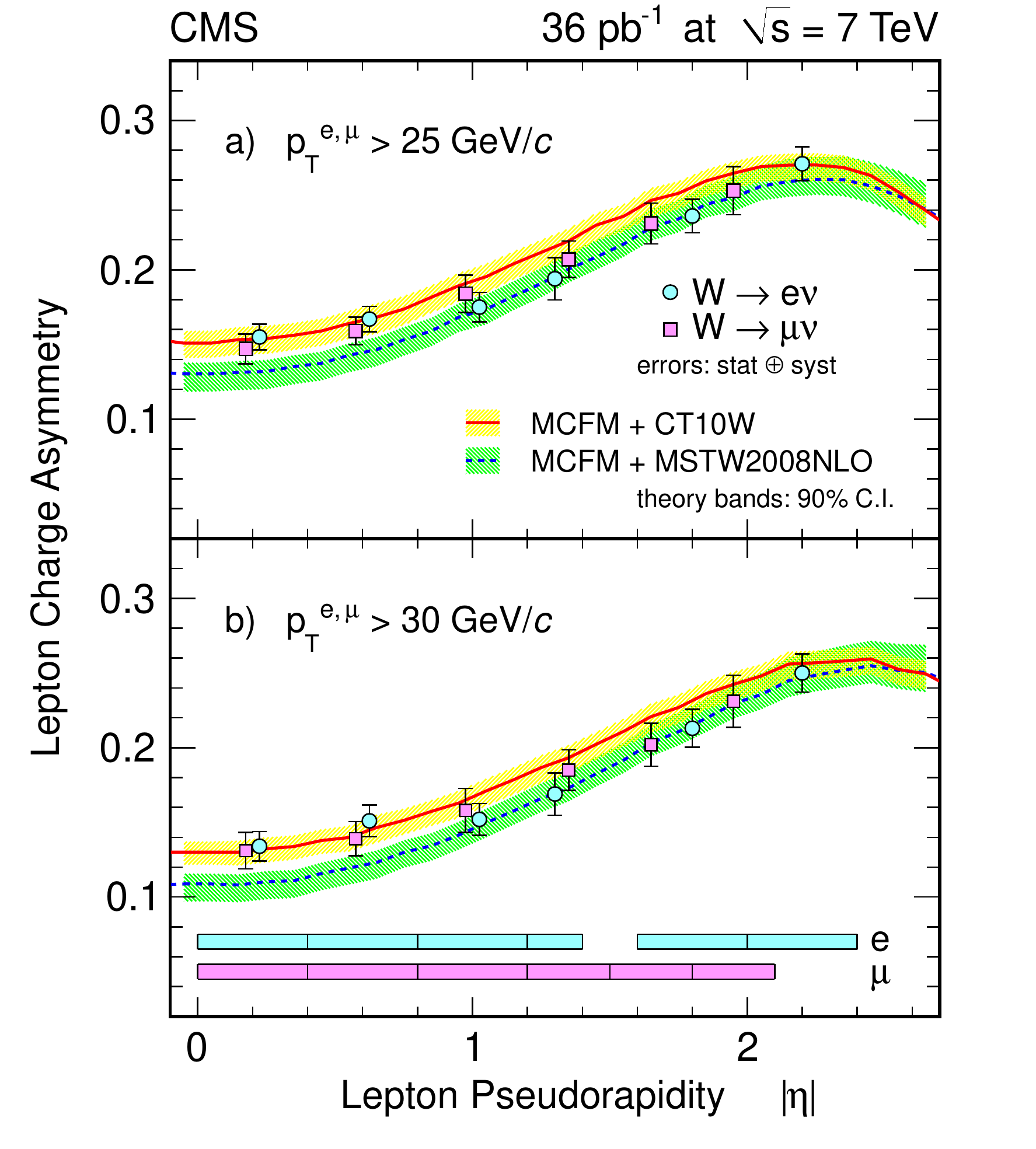}
    \hspace{1cm}
    \caption{Comparison of the measured lepton charge
asymmetry to
different PDF models for a) lepton $\pt^{\ell} >$ 25\GeVc and  b)  lepton $p^{\ell}_T >$ 30\GeVc.
The error bars include both statistical and systematic uncertainties.
The PDF uncertainty band is  corresponding to the 90\% confidence interval~(C.I.).
The bin width for each data point is shown by the filled bars in fig.  b). The data points are placed at  the centers of pseudorapidity bins,
except that for display purposes the first three data points are
shifted $+0.025$~($-0.025$) for
electron (muon). }
    \label{fig2}
  \end{center}
\end{figure}

Figure~\ref{fig2} shows a comparison of these asymmetries to
predictions from the MSTW2008NLO PDF model~\cite{MSTW:0901} and the CT10W PDF model~\cite{CTEQ:1007}. The central values of both predictions are
 obtained using the
{\sc MCFM} MC~\cite{MCFM} and the PDF error bands are estimated using
the PDF reweighting technique~\cite{PDFERROR}.
Our data suggest a flatter pseudorapidity
dependence of the
asymmetry than the PDF
models studied.
\begin{table*}[tb]
    \caption{Summary of charge asymmetry~($\mathcal{A}$) results.
The first uncertainty is
statistical and the second is systematic. The theoretical predictions are
obtained using {\sc RESBOS}~($\mathcal{A}^{\mathrm{R}}$)  and {\sc MCFM}~($\mathcal{A}^{\mathrm{M}}$) interfaced with CT10W PDF model.
 The PDF uncertainties~($\Delta (+/-)$) are estimated using the PDF reweighting
technique.
The charge asymmetries and PDF errors are given in percent.
For each pseudorapidity bin the
 theoretical prediction is  calculated using the
averaged differential cross sections for positively
and negatively charged leptons respectively. The statistical
uncertainty on the theoretical prediction is about 0.1\%. }
    \label{tab2}
    \begin{center}
{\footnotesize

      \begin{tabular*}{\textwidth}{@{\extracolsep{\fill}}  l|cccc|cccc} \hline\hline
&  \multicolumn{4}{c|}{\rule[-1.8mm]{0mm}{6.0mm} $p^{\ell}_{\mathrm{T}}>$ 25\GeVc} &
 \multicolumn{4}{c}{\rule[-1.8mm]{0mm}{6.0mm} $p^{\ell}_{\mathrm{T}} >$ 30\GeVc}\\ \hline\hline

\rule[-1.8mm]{0mm}{5.5mm}$|\eta^\textrm{e}|$&\rule[-1.8mm]{0mm}{5.5mm}  $\mathcal{A}(\textrm{e})$ ($\pm \mathrm{stat}\pm \mathrm{sys}$)&\rule[-1.8mm]{0mm}{5.5mm} $\mathcal{A}^{\mathrm{R}}$&  \rule[-1.8mm]{0mm}{5.5mm} $\mathcal{A}^{\mathrm{M}}$& \rule[-1.8mm]{0mm}{5.5mm} $\Delta (+/-)$ &\rule[-1.8mm]{0mm}{5.5mm}
 $\mathcal{A}(\textrm{e})$ ($\pm \mathrm{stat}\pm \mathrm{sys}$)&\rule[-1.8mm]{0mm}{5.5mm} $\mathcal{A}^{\mathrm{R}}$& \rule[-1.8mm]{0mm}{5.5mm}  $\mathcal{A}^{\mathrm{M}}$& \rule[-1.8mm]{0mm}{5.5mm} $\Delta (+/-)$ \\\hline

\mbox{[0.0, 0.4]} & 15.5 $\pm$ 0.6 $\pm$ 0.7 &  15.7& 15.3& ${+0.8}/{- 1.0} $&
13.4  $\pm$ 0.7  $\pm$ 0.7&  13.4& 13.1&${+0.7}/{-0.9} $\\

\mbox{[0.4, 0.8]} & 16.7   $\pm$ 0.6 $\pm$ 0.7 & 16.9&   16.7&${+0.9}/{-1.0} $&
 15.1   $\pm$ 0.7 $\pm$  0.8 &  14.6&  14.5&${+0.8}/{-0.8} $\\

\mbox{[0.8, 1.2]} & 17.5   $ \pm$ 0.7 $\pm$ 0.8 &  19.3&  19.2&${+0.8}/{- 1.1} $&
15.2 $\pm$ 0.7 $\pm$ 0.8 & 16.9&  16.8&${+0.8}/{-1.0} $\\

\mbox{[1.2, 1.4]} &  19.4  $\pm$ 1.0  $\pm$ 0.9 & 21.6&  21.7&${+0.8}/{- 1.1} $&
16.9   $\pm$ 1.1  $\pm$ 0.9 &  19.1&  18.9&${+0.8}/{-1.0} $\\

\mbox{[1.6, 2.0]} &  23.6  $\pm$ 0.8 $\pm$ 0.9 &  25.6&  25.4&${+0.8}/{- 1.1} $&
21.3  $\pm$ 0.9 $\pm$ 0.9 & 23.4&  23.7&${+0.8}/{-1.1} $\\

\mbox{[2.0, 2.4]} &  27.1 $\pm$ 0.8 $\pm$  0.9 &  27.1&  26.9&${+0.8}/{- 1.1} $&
25.0  $\pm$ 0.9  $\pm$ 0.9 & 25.7&   25.4&${+0.8}/{-1.1} $\\\hline\hline

\rule[-1.8mm]{0mm}{5.5mm} $|\eta^{\mu}|$& \rule[-1.8mm]{0mm}{5.5mm} $\mathcal{A}(\mu)$($\pm \mathrm{stat}\pm \mathrm{sys}$) &\rule[-1.8mm]{0mm}{5.5mm} $\mathcal{A}^{\mathrm{R}}$& \rule[-1.8mm]{0mm}{5.5mm}  $\mathcal{A}^{\mathrm{M}}$&\rule[-1.8mm]{0mm}{5.5mm}   $\Delta (+/-)$ & \rule[-1.8mm]{0mm}{5.5mm}
 $\mathcal{A}(\mu)$($\pm \mathrm{stat}\pm \mathrm{sys}$) & \rule[-1.8mm]{0mm}{5.5mm}  $\mathcal{A}^{\mathrm{R}}$& \rule[-1.8mm]{0mm}{5.5mm}   $\mathcal{A}^{\mathrm{M}}$& \rule[-1.8mm]{0mm}{5.5mm}  $\Delta (+/-)$ \\\hline

\mbox{[0.0, 0.4]} & 14.7 $\pm$ 0.6 $\pm$ 0.8 &  15.7&  15.3&${+0.8}/{-1.0} $&
13.1  $\pm$ 0.7  $\pm$ 1.0&  13.4&   13.1&${+0.7}/{-0.9} $\\

\mbox{[0.4, 0.8]} & 15.9   $\pm$ 0.6 $\pm$ 0.7 & 16.9&  16.7&${+0.9}/{-1.0} $&
 13.9   $\pm$ 0.7 $\pm$  0.9 &  14.6&  14.5&${+0.8}/{-0.8} $ \\

\mbox{[0.8, 1.2]} & 18.4   $ \pm$ 0.6 $\pm$ 1.1 &  19.3&  19.2&${+0.8}/{-1.1} $&
15.8 $\pm$ 0.7 $\pm$ 1.3 & 16.9&  16.8&${+0.8}/{-1.0} $\\

\mbox{[1.2, 1.5]} &  20.7  $\pm$ 0.7  $\pm$ 1.0 & 22.0&  22.0&${+0.8}/{-1.1} $&
18.5   $\pm$ 0.8  $\pm$ 1.1 &  19.6&  19.4&${+0.8}/{-1.0} $\\

\mbox{[1.5, 1.8]} &  23.1  $\pm$ 0.8 $\pm$ 1.1 &  24.6&  24.5&${+0.8}/{-1.1} $&
20.2  $\pm$ 0.8 $\pm$ 1.2 & 22.2&  21.9&${+0.8}/{-1.1} $\\

\mbox{[1.8, 2.1]} &  25.3 $\pm$ 0.8 $\pm$  1.4 & 26.5&  26.3&${+0.8}/{-1.0} $&
23.1  $\pm$ 0.9  $\pm$ 1.5 & 24.5&  24.1&${+0.8}/{-1.1} $\\

\hline\hline
      \end{tabular*}

}
    \end{center}
  \end{table*}

In summary, we have measured the lepton  charge asymmetry in both the $\wenu$ and
$\wmunu$ channels using a data sample corresponding to
 an integrated luminosity
of  $36\ \mathrm{pb}^{-1}$ collected with the CMS detector at the LHC.
In each pseudorapidity bin the precision of the most inclusive measurements
is less than 1.6\% for both channels.
This 
high precision measurement of the $\PW$ lepton charge asymmetry at the LHC
provides new inputs to
the PDF global fits.

\section{Acknowledgements}

We wish to thank James~W. Stirling for very useful discussions on PDF models and  John Campbell for providing theoretical
predictions based on MCFM.
We wish to congratulate our colleagues in the CERN accelerator departments for the excellent performance of the LHC machine. We thank the technical and administrative staff at CERN and other CMS institutes, and acknowledge support from: FMSR (Austria); FNRS and FWO (Belgium); CNPq, CAPES, FAPERJ, and FAPESP (Brazil); MES (Bulgaria); CERN; CAS, MoST, and NSFC (China); COLCIENCIAS (Colombia); MSES (Croatia); RPF (Cyprus); Academy of Sciences and NICPB (Estonia); Academy of Finland, ME, and HIP (Finland); CEA and CNRS/IN2P3 (France); BMBF, DFG, and HGF (Germany); GSRT (Greece); OTKA and NKTH (Hungary); DAE and DST (India); IPM (Iran); SFI (Ireland); INFN (Italy); NRF and WCU (Korea); LAS (Lithuania); CINVESTAV, CONACYT, SEP, and UASLP-FAI (Mexico); PAEC (Pakistan); SCSR (Poland); FCT (Portugal); JINR (Armenia, Belarus, Georgia, Ukraine, Uzbekistan); MST and MAE (Russia); MSTD (Serbia); MICINN and CPAN (Spain); Swiss Funding Agencies (Switzerland); NSC (Taipei); TUBITAK and TAEK (Turkey); STFC (United Kingdom); DOE and NSF (USA).

\bibliography{auto_generated}   

\cleardoublepage\appendix\section{The CMS Collaboration \label{app:collab}}\begin{sloppypar}\hyphenpenalty=5000\widowpenalty=500\clubpenalty=5000\textbf{Yerevan Physics Institute,  Yerevan,  Armenia}\\*[0pt]
S.~Chatrchyan, V.~Khachatryan, A.M.~Sirunyan, A.~Tumasyan
\vskip\cmsinstskip
\textbf{Institut f\"{u}r Hochenergiephysik der OeAW,  Wien,  Austria}\\*[0pt]
W.~Adam, T.~Bergauer, M.~Dragicevic, J.~Er\"{o}, C.~Fabjan, M.~Friedl, R.~Fr\"{u}hwirth, V.M.~Ghete, J.~Hammer\cmsAuthorMark{1}, S.~H\"{a}nsel, M.~Hoch, N.~H\"{o}rmann, J.~Hrubec, M.~Jeitler, G.~Kasieczka, W.~Kiesenhofer, M.~Krammer, D.~Liko, I.~Mikulec, M.~Pernicka, H.~Rohringer, R.~Sch\"{o}fbeck, J.~Strauss, F.~Teischinger, P.~Wagner, W.~Waltenberger, G.~Walzel, E.~Widl, C.-E.~Wulz
% ...
\vskip\cmsinstskip
% ...
1:~~Also at CERN, European Organization for Nuclear Research, Geneva, Switzerland\\
% ...

\end{sloppypar}
\end{document}